\documentclass[final,3p,times,twocolumn]{elsarticle}

\usepackage{graphicx}

\usepackage{mathtools}
\usepackage{color}
\usepackage{amssymb}

\newcommand{\bsat}{b_{\mathrm{sat}}}
\newcommand{\qsat}{Q_{\mathrm{sat}}}
\newcommand{\xsat}{x_{\mathrm{sat}}}

\newcommand{\xch}{x_{\mathrm{Chol}}}
\newcommand{\nbound}{b_{\mathrm{tot}}}
\newcommand{\lo}{l_{\mathrm{o}}}
\newcommand{\ldo}{l_{\mathrm{do}}}
\newcommand{\m}[2]{M_{\mathrm{{#1},{#2}}}}
\newcommand{\mself}[1]{M_{\mathrm{#1}}}
\biboptions{sort&compress,super,numbers,comma,round}

\graphicspath{{../Images/}}
\journal{BBA}
\newcommand{\nachr}{nAChR}

\begin{document}

\begin{frontmatter}

\title{Boundary lipids of the nicotinic acetylcholine receptor: spontaneous partitioning via coarse-grained molecular dynamics simulation} 

\author[ccib]{Liam Sharp} 
\author[ccib,wasu]{Reza Salari}
\author[ccib,physics]{Grace Brannigan}
\address[ccib]{Center for Computational and Integrative Biology, Rutgers University-Camden, Camden, NJ}
\address[wasu]{Now at Washington University School of Medicine in St Louis}
\address[physics]{Department of Physics, Rutgers University-Camden, Camden, NJ}

\begin{abstract}
Reconstituted nicotinic acetylcholine receptors (nAChRs) exhibit significant gain-of-function upon addition of cholesterol to reconstitution mixtures, and cholesterol affects organization of nAChRs within domain-forming membranes, but whether nAChR partitions to cholesterol-rich liquid-ordered (``raft'' or $\lo$) domains or cholesterol-poor liquid-disordered ($\ldo$) domains is unknown. We use coarse-grained molecular dynamics simulations to observe spontaneous interactions of cholesterol, saturated lipids, and polyunsaturated (PUFA) lipids with nAChRs. In binary Dipalmitoylphosphatidylcholine:Cholesterol (DPPC:CHOL) mixtures, both CHOL and DPPC acyl chains were observed spontaneously entering deep ``non-annular'' cavities in the nAChR TMD, particularly at the subunit interface and the $\beta$ subunit center, facilitated by the low amino acid density in the cryo-EM structure of nAChR in a native membrane. Cholesterol was highly enriched in the annulus around the TMD, but this effect extended over (at most) 5-10\AA. In domain-forming ternary mixtures containing PUFAs, the presence of a single receptor did not significantly affect the likelihood of domain formation.  nAChR partitioned to any cholesterol-poor $\ldo$ domain that was present, regardless of whether the $\ldo$ or $\lo$ domain lipids had PC or PE headgroups. Enrichment of PUFAs among boundary lipids was positively correlated with their propensity for demixing from cholesterol-rich phases. Long $n - 3$ chains (tested here with Docosahexaenoic Acid, DHA) were highly enriched in annular and non-annular embedded sites, partially displacing cholesterol and completely displacing DPPC, and occupying sites even deeper within the bundle. Shorter $n - 6$ chains were far less effective at displacing cholesterol from non-annular sites.
\end{abstract}


\end{frontmatter}
\section{Introduction}
\label{S:1}

The nicotinic acetylcholine receptor (nAChR) is an excitatory pentameric ligand gated ion channel (pLGIC) commonly found in the neuronal post synaptic membrane and neuromuscular junction (NMJ) in mammals 
as well as the electric organs of the \textit{Torpedo} electric ray. 
nAChRs play a fundamental role in rapid excitation within the central and peripheral nervous system, and neuronal nAChRs are also critical for cognition and memory \cite{Dani2001b, Changeux2015}. Acetylcholine is the orthosteric \nachr~ligand, but numerous other exogenous and endogenous small molecules modulate \nachr s, including nicotine, general anesthetics, the tipped-arrow poison curare,  phospholipids, cholesterol, and cholesterol-derived hormones.\cite{Klaassen2015,Taly2009}  
The larger pLGIC super family that includes \nachr s has been shown to play roles in numerous diseases related to inflammation, \cite{Patel2017,Yocum2017,Cornelison2016}, addiction \cite{Cornelison2016}, chronic pain \cite{Xiong2012}, Alzheimer's Disease \cite{Walstab2010,Picciotto_Neuroprotection_2008,MartinRuiz_4_1999}, spinal muscular atrophy \cite{Arnold_Reduced_2004}, schizophrenia \cite{Haydar2010} and neurological autoimmune diseases \cite{Lennon_Immunization_2003}.

nAChRs are highly sensitive to the surrounding lipid environment\cite{Hamouda2006a,Baenziger2017,Padilla-Morales2016,Barrantes2007} for reasons that remain poorly understood. In the late 1970s it was observed that reconstituted \nachr s only exhibit native conductance if model phospholipid membranes contained at least 10-20\% cholesterol
\cite{Dalziel1980,Criado1982,Ochoa1983}. Three generations of investigation into the mechanism have followed, with the first generation of studies\cite{Marsh1978,Dalziel1980,Marsh1981,Criado1982,Gonzalez-Ros1982,McNamee1982,Ellena1983,Ochoa1983,Zabrecky1985,Bristow1987,Leibel1987,Middlemas1987,Jones1988a,Jones1988, Fong1986,Fong1987,McNamee1988, Barrantes1989,Sunshine1992,Sunshine1994,Narayanaswami1993,Addona1998,Corbin1998,Barrantes2000} aiming to differentiate between the role of bulk, annular, and non-annular cholesterol. The second generation\cite{Baenziger2015,Bruses2001,Marchand2002,Oshikawa2003,Pato2008,Zhu2006,Baenziger2017, Barrantes2007,Barrantes2000,Barrantes2010,Bermudez2010,Perillo2016,Wenz2005,Borroni2016, Unwin2017} of studies probed membrane-mediated effects on organization of multiple \nachr s, while the third generation\cite{Basak2017,Althoff2014,Laverty2017,Zhu2018} has applied x-ray crystallography and high-resolution cryo electron microscopy to directly observe lipid binding modes. 

Members of the pLGIC family other than \nachr~are also lipid-sensitive,\cite{Dunn1989,Sooksawate2001,Baenziger2011, Dostalova2014} and lipids other than cholesterol can also modulate function\cite{Bhushan1993,Cheng2007,Corrie2002a,DaCosta2002,Rankin1997,Wenz2005,Hamouda2006a}, but these mechanisms have not been as extensively studied.  The recent publication of several crystal and cryo-EM structures \cite{Basak2017,Althoff2014,Laverty2017,Zhu2018}
has confirmed that specific lipid-pLGIC interactions extend beyond cholesterol and \nachr.  Such interactions are also well-established in other transmembrane proteins, including G-protein coupled receptors (GPCRs) and other ion channels, as reviewed in \cite{Burger2000, Lee2004, Pucadyil2006a, Landreh2016, Smithers2012}. 

Even in the specific case of cholesterol-\nachr~ interactions, results from different approaches have suggested complex behavior and even contradictory interpretations.  Results have indicated that both cholesterol enrichment\cite{Dalziel1980,Criado1982,Ochoa1983} and cholesterol depletion\cite{Santiago2001} cause gain of function, that anionic phospholipids are unnecessary for native function\cite{Dalziel1980,Criado1982,Ochoa1983} or must be\cite{Corrie2002a,DaCosta2002} included in a reconstitution mixture,  that cholesterol increases \nachr~clustering\cite{Pato2008, Zhu2006, Barrantes2007} and directly interacts with \nachr~\cite{Leibel1987,Jones1988}, but \nachr~does not consistently partition into cholesterol-rich domains\cite{Bermdez_Partition_2010}. We suggest here that some of these apparent contradictions may be explained by competition between cholesterol and other lipids found in native membranes, primarily lipids with polyunsaturated fatty acyl chains (PUFAs).   

Interactions of \nachr~ with PUFAs have not been systematically investigated experimentally, but a large amount of circumstantial experimental evidence suggests an important role for PUFAs in \nachr~ function.    Clinically, long-chain $n-3$  (commonly called ``Omega-3'' or $\omega-3$) lipids have a neuroprotective role\cite{Piomelli2007}, and \nachr-associated pathologies can arise for patients with low levels of $n-3$ PUFAs. $\alpha7$ \nachr s are implicated in schizophrenia\cite{Haydar2010}, and dietary supplementation with $n-3$ fatty acids (usually through fish oil) can reduce the likelihood of psychosis, with dramatic effects in some individual cases.\cite{Amminger2010}   

{\it In vitro}, PUFA-rich asolectin\cite{Regost2003,Olsen2003} is one of the most robust additives\cite{Criado1982} for obtaining native \nachr~function: restoration of native function by cholesteryl hemisuccinate (CHS) is observed only over a narrow CHS concentration range in monounsaturated PE/PS membranes, but a much wider concentration range in asolectin\cite{Criado1982}. The specific component(s) of asolectin that complement cholesterol in improving \nachr~function have not been isolated. 
Long chained $n-3$  
PUFA lipids are abundant in two seemingly disparate \nachr~native membranes: mammalian neuronal membranes\cite{Breckenridge_Adult_1973,Cotman_Lipid_1969} and those of the \textit{Torpedo} electric organ,\cite{Barrantes1989,Quesada2016}. Both such membranes also have an abundance of phosphoethanolamine (PE) headgroups and saturated glycerophospholipids, and a scarcity of monounsaturated acyl chains and sphingomyelin compared to thhe \textit{Xenopus} oocyte membranes \cite{Hill_Isolation_2005} common in functional studies, or a ``generalized'' mammalian cell membrane \cite{Inglfsson_Lipid_2014}.    

Membranes composed of ternary mixtures of saturated lipids, unsaturated lipids, and cholesterol tend to demix into separate domains. Saturated lipids and cholesterol constitute a rigid liquid ordered phase ($\lo$) in which acyl chains remain relatively straight. \cite{Feller_Acyl_2008,Yeagle2016115,Cicuta1981,Bleecker2016} Unsaturated lipids form a more flexible liquid disordered phase ($\ldo$) in which the chains remain fully melted.  
$\lo$ domains are often visualized as signaling ``platforms'', restricting membrane proteins into high density ``rafts'' that diffuse within a fluid membrane {\cite{Simons1997,Simons2000}}. This conceptualization requires that $\lo$ domains have a much smaller area than $\ldo$ domains, and does not well-represent membranes that are over 30\% cholesterol, such as neuronal membranes.   

The first generation of studies into the mechanism underlying cholesterol-modulation of \nachr~ were conducted and interpreted in an era preceding the discovery of lipid-induced domain formation in membranes. 
The second generation explicitly considered potential interactions of \nachr~with lipid domains, in part to determine the requirements for the extremely high density  ($\sim10^{4}\mu^{-2}$) of \nachr s at the neuromuscular junction \cite{Breckenrldge1972}.  Since direct interaction between \nachr~and cholesterol had been demonstrated in the first generation of studies, a sensible initial hypothesis was that \nachr~persistently partitioned to $\lo$ domains, retaining little contact with unsaturated chains.  Tests of this hypothesis have yielded results that are inconclusive, contradictory, or highly sensitive to lipid composition.     

Barrantes and colleagues\cite{Wenz2005} found that the addition of \nachr s~to a domain-forming lipid mixture increased the size of Dipalmitoylphosphatidylcholine/Cholesterol (DPPC/Chol) lipid-ordered domains, which (combined with additional FRET data) was interpreted as indicating \nachr~was embedded in liquid-ordered domains.  Some studies \cite{Marchand2002,Stetzkowski-Marden2006,Willmann2006} suggest that \nachr s are associated with microdomains independently of stimulation by other proteins associated with the neuromuscular junction. Other studies\cite{Zhu2006,Campagna2006} suggested that \nachr s require stimulation by a protein such as agrin to partition into microdomains. Formation and disassembly of the \nachr-rich microdomains is highly sensitive to cholesterol concentration.\cite{Barrantes2007,Bruses2001,Marchand2002,Zhu2006,Pun2002}

These studies suggested a role for cholesterol-induced phase separation, but did not confirm that \nachr~partitions to the cholesterol-rich phase.  To test for an intrinsic \nachr~ domain preference, Barrantes and co-workers checked for enrichment of \nachr s in the detergent resistant membrane (DRM).   \nachr s were not enriched in the DRM of a model, domain-forming mixture (1:1:1  Chol: palmitoyloleoylphosphatidylcholine(POPC): sphingomyelin) \cite{Bermdez_Partition_2010} but inducing compositional asymmetry across leaflets did yield \nachr~enrichment in the DRM fraction \cite{Perillo_Transbilayer_2016}.  While more precise and robust experimental methods for determining partitioning preference and specific boundary lipids such as mass spectrometry have been applied for other transmembrane proteins\cite{Gupta2018,Chorev2018}, they have not been applied to complex heteromers like \nachr.  

Fully atomistic molecular dynamics (MD) simulations\cite{Brannigan_Embedded_2008, Cheng2009, Hnin_A_2014, Carswell_Role_2015} have served as a natural complement to the third-generation structural biology approach, but are limited in their ability to resolve contradictions between first and second generation studies, because lipids are unable to diffuse over simulation time scales.\cite{Ingolfsson2014,Bond2006,Parton2013,Goose2013,Scott2008}.   Efficient lipid diffusion is a requirement for equilibrating domains or detecting protein-induced lipid sorting.    Coarse-grained MD (CG-MD) has been used to great success in a number of simulations for both lipid-protein binding and membrane organization \cite{Bond2006,Scott2008,Parton2013,Goose2013,Iyer2018,Sodt2014}. Here we use CG-MD as a ``computational microscope'' to observe the equilibrium distribution of lipids local to the \nachr~ in a range of binary and ternary lipid mixtures inspired by native membranes.   We observe a remarkable enrichment of polyunsaturated lipids among \nachr~boundary lipids. To our knowledge, these are the first molecular simulations of the \nachr~in non-randomly mixed membranes, and the first study to systematically investigate the likelihood of polyunsaturated lipids as \nachr~boundary lipids.  \section{Methods}
\label{S:2}

\subsection{System Composition}

All simulations reported here used the coarse-grained MARTINI 2.2\cite{martini} topology and forcefield.
~nAChR coordinates were based on a cryo-EM structure of the $\alpha{\beta}\gamma\delta$ muscle-type receptor in native torpedo membrane (PDB 2BG9\cite{Unwin_Refined_2005}). This is a medium resolution structure (4\AA) and was further coarse-grained using the martinize.py script; medium resolution is sufficient for use in coarse-grained simulation, and the native lipid environment of the proteins used to construct 2BG9 is critical for the present study. The secondary, tertiary and quaternary structure in 2BG9 was preserved via soft backbone restraints during simulation as described below, so any inaccuracies in local residue-residue interactions would not cause instability in the global conformation.  

Coarse-grained membranes were built using the Martini script insane.py, which was also used to embed the coarse-grained \nachr~within the membrane. The insane.py script randomly places lipids throughout the inter- and extra-cellular leaflets, and each simulation presented in this manuscript was built separately.  Binary mixed membranes were composed of one saturated lipid species (Dipalmitoylphosphatidylcholine-DPPC or Dipalmitoylphosphatidylethanolamine-DPPE) and cholesterol (CHOL), while ternary mixed membranes also included either two $n-6$ PUFA acyl chains : Dilinoleoylphosphatidylcholine (dLA-PC) or Dilinoleoylphosphatidylethanolamine (dLA-PE) or two $n-3$ PUFA acyl chains : Didocosahexaenoylphosphatidylethanolamine (dDHA-PE) or Didocosahexaenoylphosphatidylcholine (dDHA-PC). DHA-PC is not distributed with the MARTINI lipidome, but was constructed in-house using MARTINI DHA tails and PC headgroups). Multiple box sizes were used depending on the goal;  ``small'' boxes were between $22x22x20$ nm$^3$ and $25x25x25$ nm$^3$, with about {$\sim$ 1400} total lipids and {$\sim$ 80000} total beads, and were used primarily to investigate composition trends, ``large'' boxes were about $45x45x40$ nm$^3$ with about {$\sim$ 8,300} total lipids and {$\sim$ 820,000} total beads, and were used primarily to investigate subunit specificity and long-range sorting, and ``very large'' boxes were $\sim$ 75x75x40~nm$^3$ with about {$\sim$ 19,000} total lipids and {$\sim$ 1.8 million} total beads, and were used to verify that partitioning in the $\ldo$ phase did not reflect finite size effects.

\subsection{Simulations}

Molecular dynamics simulations were carried out using GROMACS\cite{grom}; small boxes used GROMACS 5.0.6 and large {and very large} boxes used  GROMACS 5.1.2 or 5.1.4. All systems were run using van der Waals (vdW) and Electrostatics in shifted form with a dielectric constant of $\epsilon_r$=15. vdW cutoff lengths were between 0.9 and 1.2 nm, with electrostatic cutoff length at 1.2 nm.

Energy minimization was performed over 10000 to 21000 steps.  Molecular dynamics were run using a time step of 25~fs, as recommended by MARTINI, for 2 $\mu$s for {small membranes,and 10 $\mu$s for large and very large membranes}. Simulations were conducted in the isothermal-isobaric (NPT) ensemble, by using a Berendsen thermostat set to 323 K with temperature coupling constant set to  1 ps, as well as isotropic pressure coupling with compressibility set to $3\times 10^{-5}$ bar$^{-1}$ and a pressure coupling constant set to 3.0 ps.

Secondary structures restraints consistent with MARTINI recommendations were constructed by the martinize.py \cite{martini} script {and} imposed by Gromacs\cite{grom}. Protein conformation was maintained in small systems via harmonic restraint (with a spring constant of 1000 kJ$\cdot$ mol$^{-1}$) on the position of backbone beads. \nachr~conformation in large systems was preserved via harmonic bonds between backbone beads separated by less than 0.5 nm, calculated using the ElNeDyn algorithm \cite{Periole_Combining_2009} associated with MARTINI\cite{martini}  with a coefficient of 900 kJ$\cdot$ mol$^{-1}$.  These restraints limited the root-mean-squared-displacement (RMSD) of the backbone to less than 2.5 \r{A} throughout the simulation.  

The minimum equilibration time depended on the system size. Small systems typically began domain formation by 500 ns, with domains fully formed by 1000 ns. Large systems and very large simulations required about 5$\mu$s of equilibration for stabilization of metrics described below.

\subsection{Analysis}

Extent of domain formation within the membrane was tracked by 
    \begin{equation}
    \begin{aligned}
      \m{A}{B} &\equiv \frac{\langle n_{A,B} \rangle} {6x_{B}} -1 \\
      \mself{A} &\equiv \frac{\langle n_{A,A} \rangle} {6x_{A}} -1 
    \end{aligned}
    \label{eq:M}
  \end{equation}
 where $n_{A,B}$ is the number of type B molecules among the 6 nearest neighbors for a given type A molecule,  the average is over time and all molecules of type $A$, and the self-association metric is notated $\mself{A} \equiv \m{A}{A}$ for brevity. For a random mixture, $\langle n_{A,B} \rangle = 6x_{B}$, where $x_{B}$ is the fraction of overall bulk lipids that are of type B. ${\mself{A}=0}$ indicates random mixing while ${\mself{A}>0}$  and ${\mself{A}<0}$ indicate demixing and excessive mixing respectively.

Extent of receptor partitioning within the $\lo$ or $\ldo$ domain was tracked by counting the number $\bsat$ of saturated annular boundary lipids and comparing with the expectation for a random mixture, via the order parameter $\qsat$:
  \begin{equation}
    \begin{aligned}
      \qsat\equiv \frac{1}{\xsat}\left\langle\frac{  \bsat }{\nbound }\right\rangle-1,\\
    \end{aligned}
    \label{eq:Q}
  \end{equation}
  where $\nbound$ is the total number of lipids in the annular boundary region and $\xsat$ is the fraction of overall bulk lipids that are saturated phospholipids. $\qsat <0$ indicates depletion of saturated lipids among boundary lipids, as expected for partitioning into an $\ldo$ phase, while $\qsat>0$ indicates enrichment and likely partitioning into an $\lo$ phase. Each frame, $\nbound$ and $\bsat$ were calculated by counting the number of total and saturated lipids, respectively, for which the phosphate bead fell within a distance of 1.0~nm~ to 3.5~nm~ from the M2 helices, projected onto the membrane plane. 
  
  Two-dimensional density distribution of the beads within a given lipid species $B$ around the protein was calculated on a polar grid: 
  \begin{equation}
    \begin{aligned}
      \rho_{B}(r_i,\theta_j)= \frac{\left\langle n_{B}(r_i,\theta_j) \right\rangle}{r_i \Delta{r}\Delta{\theta}} \\        
    \end{aligned}
    \label{eq:R}
  \end{equation}
  where  $r_i = i \Delta{r}$ is the projected distance of the bin center from the protein center, $\theta_j = j \Delta{\theta}$ is the polar angle associated with bin j,  $\Delta{r}$= 10\AA~ and  $\Delta{\theta} = \frac{\pi}{15}$ radians are the bin widths in the radial and angular direction respectively, and $\left\langle n_{B}(r_i,\theta_j) \right\rangle$ is the time-averaged number of beads of lipid species $B$ found within the bin centered around radius $r_{i}$ and polar angle $\theta_{j}$.  In order to determine enrichment or depletion, the normalized density $ \tilde{\rho}_{B}(r_i,\theta_j)$ is calculated by dividing by the approximate expected density of beads of lipid type B in a random mixture, $x_{B}s_{B}~N_{L}/\langle L^{2}\rangle$, where $s_{B}$ is the number of beads in one lipid of species B, $N_{L}$ is the total number of lipids in the system, and $\langle L^{2}\rangle$ is the average projected box area: 
  \begin{equation}
    \begin{aligned}
  \tilde{\rho}_{B}(r_i,\theta_j)=\frac{ \rho_{B}(r_i,\theta_j)}{x_{B}s_{B}~N_{L}/\langle L^{2}\rangle} \\        
    \end{aligned}
    \label{eq:Rt}
  \end{equation}
  This expression is approximate because it does not correct for the protein footprint or any undulation-induced deviations of the membrane area.  The associated corrections are small compared to the membrane area and would shift the expected density for all species equally, without affecting the comparisons we perform here.

   \section{Results}
\label{S:3}
\begin{figure*}[h!]
	\center
	\includegraphics[width=\linewidth]{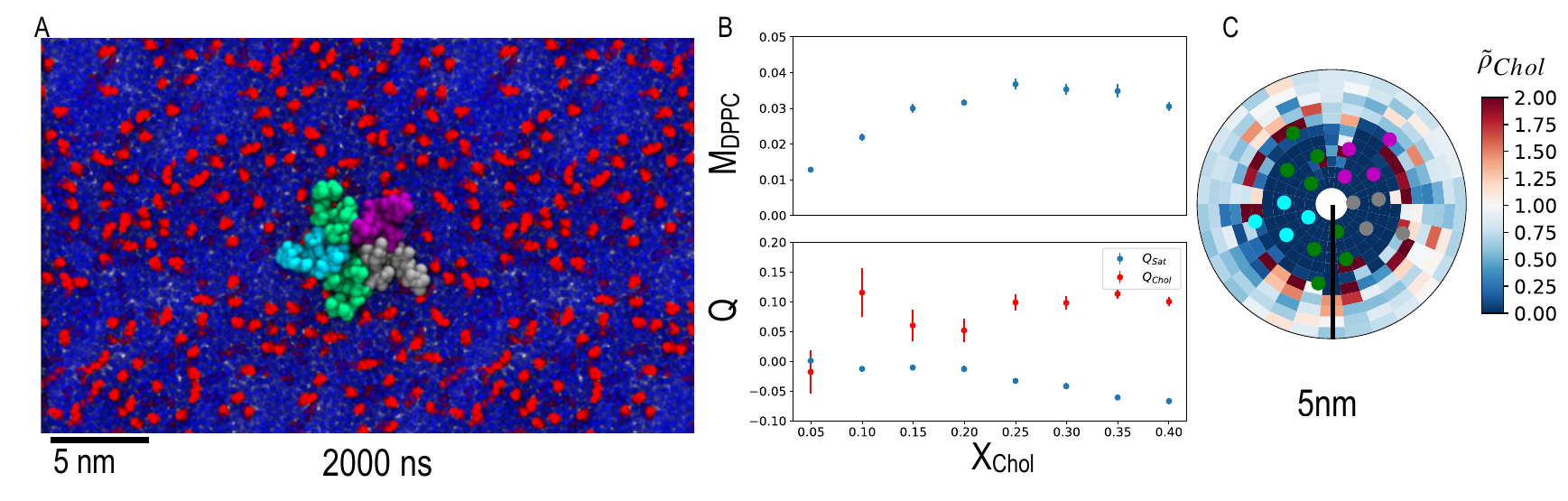}
	\caption{nAChR boundary lipids in binary mixtures of DPPC and CHOL. A: Representative frame from a simulated trajectory of a single nAChR embedded in a small membrane, colored by subunit ($\alpha$:green, $\beta$:purple, $\delta$:gray, $\gamma$:cyan) in a 4:1 DPPC (blue):Chol (red) mixture.  B: Extent of demixing ($\mself{DPPC}$ defined in Eq. \ref{eq:M}) and depletion of saturated lipids from the boundary ($\qsat$ defined in Eq.\ref{eq:Q}) in small binary membranes. In this binary system, cholesterol depletion/enrichment is directly related to the saturated lipid depletion/enrichment: $Q_\mathrm{chol}=-x_{\mathrm{sat}} \qsat/\xch$.  Error bars represent standard error for a blocking average over 50 ns. C: Average normalized density (Eq. \ref{eq:Rt}) of cholesterol for the system in A. Data is equivalent to that in Figure \ref{fig:sorting}: Binary Mixture ``Chol'' row.}
	\label{fig:binary}
\end{figure*} 
\subsection {Spontaneous association with cholesterol in binary membranes} \label{binary}

Lipid sorting was characterized for \nachr s in binary DPPC:CHOL membranes (Figure \ref{fig:binary}A)  using several metrics. 
Non-random lipid mixing (including domain formation) was quantified using the self-association metric $\mself{A}$ as defined in Equation \ref{eq:M}. 
As expected, in simulated binary membranes containing only DPPC and 0-40\% cholesterol, minimal demixing was observed, with values of $\mself{DPPC}$ (Fig \ref{fig:binary}B) rising slightly for higher cholesterol concentrations but remaining persistently below 0.05.  

Depletion of saturated lipids among \nachr~boundary lipids (relative to those expected for a random mixture) was quantified using the metric $\qsat$ defined in equation \ref{eq:Q}. Negative and positive values of $\qsat$ reflect depletion or enrichment of saturated lipids in the \nachr~boundary, respectively. In binary systems containing cholesterol and saturated lipids, depletion of saturated lipids corresponds directly to enrichment of cholesterol: $Q_{chol} = -\qsat x_\mathrm{sat}/\xch$. 

In binary DPPC:CHOL mixtures, $\qsat$ was very slightly negative for $\xch < 20\%$, but decreased steadily for higher concentrations. This trend indicates some depletion of DPPC (and enrichment of cholesterol) among \nachr~ boundary lipids (Figure \ref{fig:binary}B).  Typically, between 10 and 20\% cholesterol has been required in reconstitution mixtures to restore native function  \cite{Fong_Correlation_1986,Dalziel1980,Criado1982}  and a phase transition at about 20\% cholesterol in binary DPPC:CHOL model membranes is indicated by differential scanning calorimetry.\cite{Marsh2010} 

Spontaneous binding of cholesterol to non-annular or ``embedded'' sites, similar to what we previously proposed\cite{Brannigan_Embedded_2008}, was observed in these CG-simulations, and penetration of the TMD bundle by DPPC acyl chains was also observed at lower cholesterol concentrations (Fig \ref{fig:binary}A).  Distribution of density for embedded lipids is further discussed in Section 3.4.  

Annular cholesterol (enrichment of cholesterol at the protein-lipid interface), is visible for the binary systems via a ring of high (red) cholesterol density just around the protein in Figure \ref{fig:binary}C. Enrichment of cholesterol near the protein is highly localized with a ring that is less than 5\AA~wide. This is in general agreement with evidence for annular cholesterol in randomly-mixed binary membranes. \cite{Barrantes2010}

\subsection {Domains formed in PUFA-containing ternary membranes are not affected by introduction of an \nachr } \label{Demix}
	\begin{figure}[h!]
		\center
		\includegraphics[width=1\linewidth]{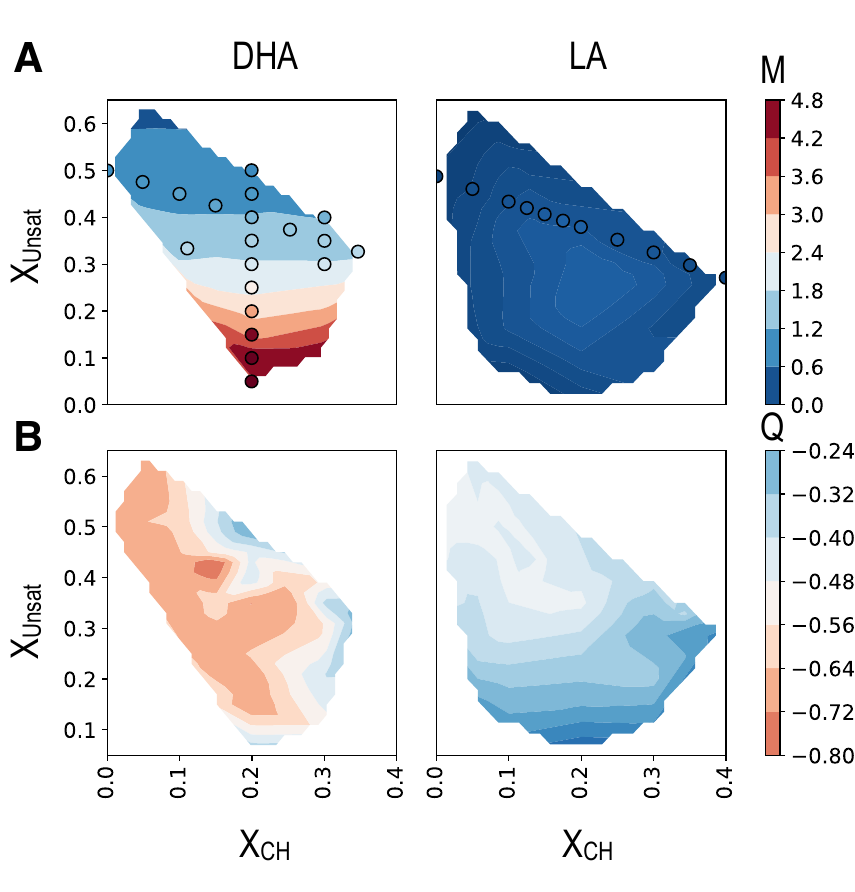}
		\caption{Quantitative analysis of bulk membrane mixing and nAChR boundary lipid composition across small membranes containing DPPC, Cholesterol, and either dDHA-PE or dLA-PC. Shaded contours were constructed based on 40 individual simulations with dDHA-PE and 30 with dLA-PC. A: $\mself{PUFA}$, defined in eq \ref{eq:M}.  Circles represent mixing of systems with the same lipid composition but no \nachr. B: $\qsat$, defined in Eq \ref{eq:Q}.   }
		\label{fig:fig2}
	\end{figure} 
	
	In order to test whether \nachr~affected domain formation in domain-forming membranes, we characterized $\mself{PUFA}$ for systems containing DPPC, Cholesterol, and PE or PC with either n-3 (DHA) or n-6 (LA) acyl chains.   Addition of phospholipids with unsaturated acyl chains to systems containing a saturated lipid and cholesterol is well-established to induce domain formation, and polyunsaturated phospholipids make these domains more well-defined\cite{Levental_Polyunsaturated_2016}. As expected, we observed that addition of PUFAs to DPPC/CHOL bilayers did induce domain formation over a range of compositions, and values for $\mself{PUFA}$ are shown as filled symbols in Figure \ref{fig:fig2} A. 
	
\begin{figure*}[t]
		\center
		\includegraphics[width=1\linewidth]{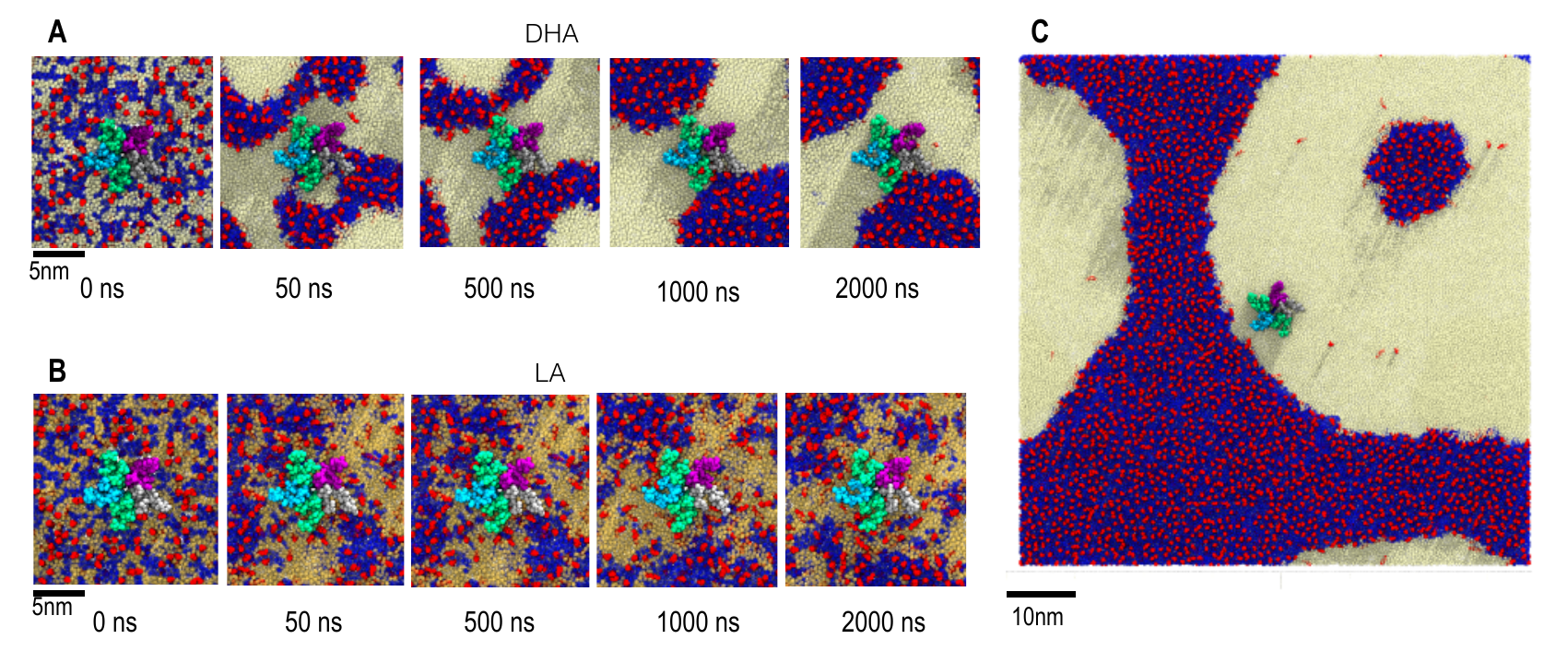}
		\caption{ Trajectories of ternary mixtures at ratios of 2:2:1 DPPC:PUFA:Chol. A and B: Trajectories of simulation systems with a single nAChR embedded within small membranes, using lipids containing DHA acyl chains or LA acyl chains. Both simulations were run for 2 $\mu$s. C: Final snapshot of 4 $\mu$s trajectory of a system within a large $\sim$ 75x75 nm$^2$ membrane with the same composition as in A. Subunits are colored: $\alpha$: green, $\beta$: purple, $\delta$: gray, $\gamma$: cyan. Lipids are colored: Chol: red, DPPC: blue, dDHA-PE: white, dLA-PC: tan.} 
		\label{fig:fig1}
	\end{figure*}

	Introducing a single nAChR to these same systems did not significantly affect domain formation. $\mself{DHA}$ was determined for an isolated \nachr~in ternary mixed membranes with over 40 different combinations of DHA, DPPC, and Cholesterol (Figure \ref{fig:fig2}A, shaded contours). Its effect on membrane organization is represented by the difference in color of the circular symbol and the shaded contour at the same composition.  Introducing a single nAChR into the DHA-containing systems does slightly reduce the amount of DHA required to obtain a given value of $\mself{DHA}$. 
	{ This subtle trend} may reflect increased likelihood of DHA-DHA interactions due to nucleation of DHA-containing lipids around the protein (Figure \ref{fig:fig1}). 

	\begin{figure}[!ht]
		\center
		\includegraphics[width=1\linewidth]{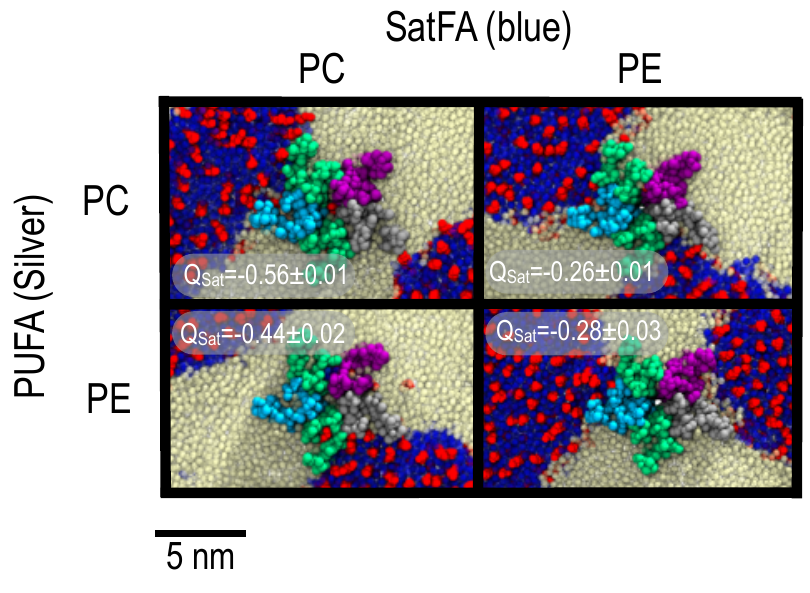}
		\caption{ Comparison of nAChR partitioning based on lipid headgroups (PC and PE). All images represent last frame of 2$\mu$s  simulations of small membranes with composition  2:2:1 Sat:PUFA:Cholesterol.  Rows represent the head-group for the PUFA-containing lipid, while columns represent the head-group of the saturate lipid.   Each image includes $\qsat$ values related to individual systems with errors across averaging 50~ns blocks.}
		\label{fig:SIQ}
	\end{figure}
	Across ternary mixtures with two long $n-3$ PUFA chains (DHA) and a PE headgroup, maximum values of $\mself{DHA}$ approached 5 (Figure \ref{fig:fig2}A), and were significantly reduced (to less than 0.5) when DHA chains were replaced with linoleic acyl (LA) chains. This result is consistent with a previously-observed significant increase in miscibility temperature upon supplementation of plasma membranes with $n-3$ lipids.  \cite{Levental_Polyunsaturated_2016} 
	
	Substantial lipid demixing in DHA-containing mixtures was observed even at low cholesterol concentrations. Over the range we tested, $\mself{DHA}$ was not sensitive to cholesterol concentration $\xch$, as shown by the horizontal contours for DHA in Figure \ref{fig:fig2}A.   

\subsection{nAChR consistently partitions to the liquid disordered domain} \label{PUFA}
	For more than 70 lipid compositions tested, nAChR always partitioned into a PUFA-rich $\ldo$ phase if such a phase was present. We never observed \nachr~partitioning to an $\lo$ phase. Representative frames from trajectories of domain formation in the presence of \nachr~are shown in Figure \ref{fig:fig1}.  This observation includes all tested concentrations of the ternary mixtures, regardless of whether the zwitterionic headgroup was PC or PE (Figure \ref{fig:SIQ}), or whether DPPC was replaced by dioleoylphosphatidylcholine (DOPC) (di-18:1), Palmitoyloleoylphosphatidylcholine (POPC) (16:0,18:1), or dilauroylphosphatidylcholine (DLPC) (di-14:0), as shown in Figure S1.   
	
	{These results are quantified for \nachr~embedded in ternary membranes containing DPPC, CHOL, and either dDHA-PE or dLA-PC in Figure \ref{fig:fig2} B, using the metric $\qsat$ defined in equation \ref{eq:Q}.}  
	In all systems studied here, $\qsat < 0$, indicating depletion of saturated lipids as boundary lipids, consistent with observed partitioning to the $\ldo$ domain in Figure \ref{fig:fig1}. Furthermore, depletion was much stronger in systems containing DHA ($\qsat^{DHA}<< \qsat^{LA}$), consistent with the more well-defined DHA domains ($\mself{DHA}>> \mself{LA}$). 

	\begin{figure}[ht!]
		\center
		\includegraphics[width=.75\linewidth]{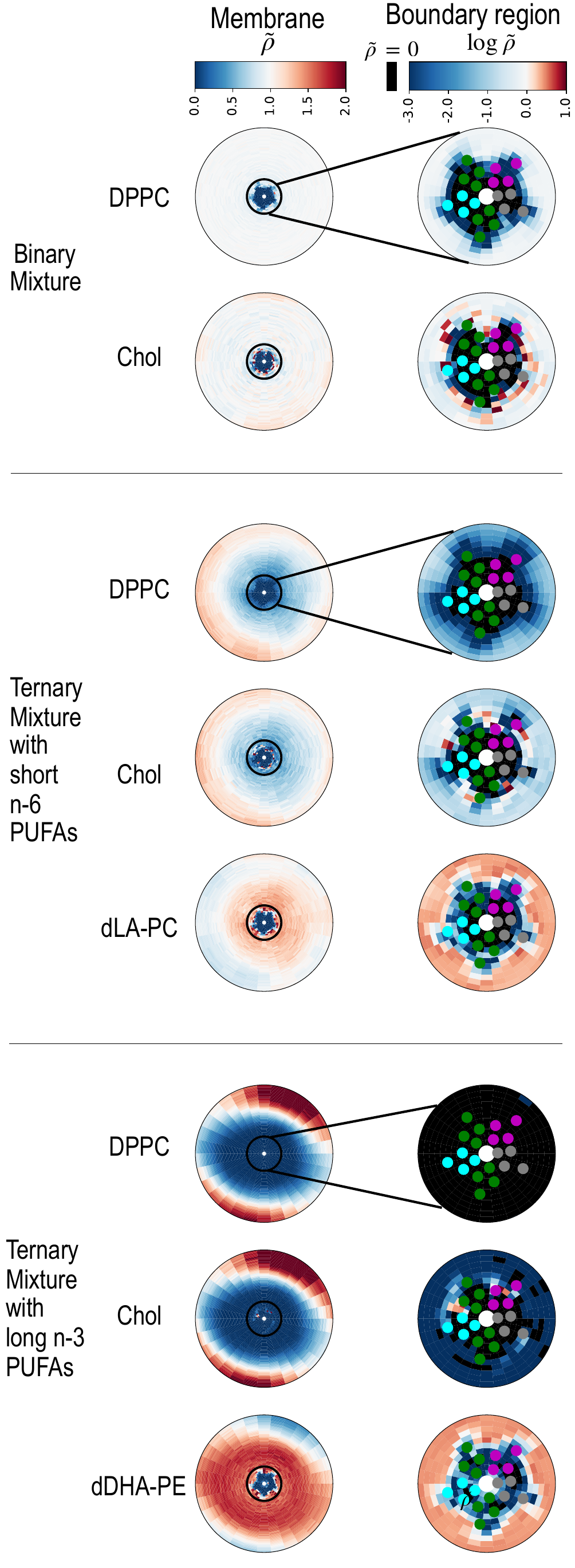}
		\caption{Lipid density enrichment or depletion around a single central nAChR.  Heatmaps are colored according to the normalized density $\tilde{\rho}_{a}$ (left, defined in eq \ref{eq:Rt}) or $\ln\tilde{\rho}_{a}$ (right), averaged over the final 5$\mu s$ of a 10 $\mu s$ simulation. Membrane column (left) depicts density across the simulated membrane; $\tilde{\rho}_{a}<1$ indicates depletion compared to a random mixture, while $\tilde{\rho}_{a}>1$ indicates enrichment. Boundary column (right) shows a zoomed-in region around the protein, with circles corresponding to average position of the protein helices, colored as in Figure 1, and black indicating no detected lipid density. If no non-annular or embedded lipid binding was observed, the entire protein footprint would be black for all lipids. Binary mixture contains 4:1 DPPC:CHOL as in Figure 1, while both ternary mixtures contain 2:2:1 DPPC:PUFA:Chol. }
		\label{fig:sorting}
	\end{figure}
	
	The \nachr~annulus is highly enriched in DHA: DHA-PE constitute nearly 100\% of the local lipids even in membranes with very low DHA concentrations. This strong signal could indicate multiple high affinity sites for DHA chains across the transmembrane protein surface. At another extreme, DHA enrichment could be driven by a very slight preference for DHA in a highly non-ideal bulk: since DHA is found in well-defined domains without protein, even one DHA molecule that binds to the protein surface could stabilize the rest of the $\ldo$ domain nearby. Comparing boundary lipid and domain formation trends can help distinguish between these two scenarios.  If boundary lipid enrichment is determined purely by how well-defined domains are (the latter scenario), we would expect similar trends for $\mself{DHA}$ and $Q_{sat}$ in the DHA column of Figure \ref{fig:fig2}.  In contrast,  Figure \ref{fig:fig2} shows that while domain formation in DHA-containing systems is only weakly sensitive to cholesterol content (horizontal contours), composition of boundary lipids is highly sensitive to cholesterol content (diagonal contours). These results suggest that direct interactions between multiple favorable sites on \nachr and DHA-containing lipids dominate the observed enrichment of DHA among boundary lipids.  
	
	The simulations represented in Figure \ref{fig:fig2} do compare the effects of two unsaturated lipids that also have different headgroups. DHA is far more commonly paired with PE in native membranes, while LA is more commonly found with PC. We found no qualitative differences in \nachr~domain partitioning or significant quantitative effect on $\qsat$ upon switching PC and PE headgroups on the PUFA lipid.  We did observe a quantitative effect of \emph{saturated} lipid headgroup on boundary lipid composition: $\qsat$ was reduced by half when saturated PE was used instead of saturated PC. (Figure \ref{fig:SIQ}).  As shown in Figure \ref{fig:SIQ}, \nachr~is bordered by $\lo$ domains on two opposing faces when saturated PE is used, compared to only one face if PC is used.  
The particular domain topology shown in Figure \ref{fig:SIQ} is an artifact of the periodic boundary conditions, but still indicates more favorable interactions of \nachr~with an $\lo$ domain composed of DPPE vs DPPC. This may reflect a difference in the lipid shape (wedge-shaped DPPE vs cylindrical-shaped DPPC) and the associated monolayer spontaneous curvature.  For PUFA lipids in flexible $\ldo$ domains, lipid shape is less likely to play a significant role in determining partitioning. The dramatic difference in domain flexibility is apparent in  Figure S2.   
	
	\subsection{Spontaneous integration of lipids into nAChR TMD bundle} \label{Embed}

	The \nachr~structure used for these simulations was determined in a native membrane with a high fraction of polyunsaturated lipids. While we previously \cite{Brannigan_Embedded_2008} proposed that unresolved density in this structure could be embedded cholesterol, the possibility of occupation by phospholipids other than POPC was not investigated.  Furthermore, we did not consider possible asymmetry across subunits in binding previously.  Here we do observe penetration of both the intersubunit (``type B'') and the intrasubunit (``type A/C'') sites previously proposed\cite{Brannigan_Embedded_2008}, by both phospholipids and cholesterol, but with a high degree of subunit specificity.  
		
\begin{figure}[h!]
		\center
		\includegraphics[width=1\linewidth]{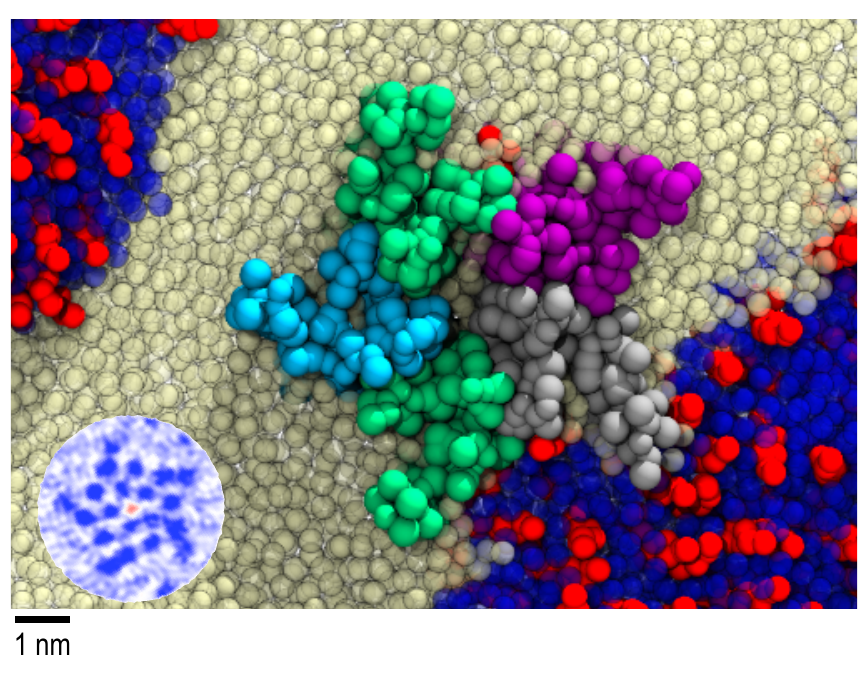}
		\caption{ Embedded lipids in the nAChR. Main image: Representative frame from equilibrated small membrane simulation of nAChR in 2:2:1 DPPC:DHA-PE:CHOL. Backbone beads of the TMD helices are colored by subunit as in Figure \ref{fig:fig2}; side-chain beads are not shown.  Both DHA-PE (white) and cholesterol (red) equilibrate to embedded sites in the subunit center and subunit interfaces, although most cholesterol is found in the $\lo$ phase with DPPC (blue). Inset : Cryo-EM density of nAChR from \cite{Miyazawa2003} as rendered in \cite{Brannigan_Embedded_2008}; dark blue indicates high density, white is medium density, and red is low density. } 
		\label{fig:sum}
	\end{figure}

Two dimensional density distributions of DPPC, PUFAs, and cholesterol over short and long length scales were measured for two ternary mixtures and one binary mixture (Figure \ref{fig:sorting}).   In binary DPPC/cholesterol membranes, DPPC was more likely than cholesterol to occupy intrasubunit sites.  DPPC binds shallowly in the $\alpha$ subunit and more deeply in the $\beta$ subunit. Introducing PUFAs resulted in displacement of both cholesterol and DPPC from intrasubunit sites, except for the $\beta$ intrasubunit site, which became more likely to be occupied by cholesterol. The interior of the $\beta$ subunit TMD has the largest amount of available volume, could sequester cholesterol (but not DPPC) from the PUFA lipids in the annulus, and filling the interior with a PUFA chain may be entropically costly.  PUFA chains did occupy other intrasubunit sites, but remained fluid, as shown in Figure \ref{fig:sum}. 

	Intersubunit sites were rarely occupied by DPPC, with the exception of the $\beta+/\alpha-$ site in the binary system (Figure \ref{fig:sorting}). Intersubunit sites were more likely to bind cholesterol, particularly the $\beta+/\alpha-$, $\alpha+/\gamma-$, and $\alpha+/\delta-$ subunit interfaces. Occupation of the $\alpha+/\delta-$ interface is consistent with cryo-EM observations\cite{Unwin_Segregation_2017}  of enhanced cholesterol density around the $\alpha+/\delta-$ site. Intersubunit sites that were not significantly occupied by cholesterol ($\delta+/\beta-$ and $\gamma-/\alpha+$) did show significant and deep occupation by DHA, which tended to enter from the adjacent intrasubunit site rather than from the membrane. Even those intersubunit sites with significant cholesterol occupancy can simultaneously bind part of a DHA chain, yielding non-vanishing DHA density.  

	\subsection{Lipid sorting over the 5-20 nm range is associated with larger domains  } \label{Sorting}

	We also calculated density distributions of each lipid species at distances beyond the ``annular'' ring, over the 5-20 nm range.    As shown in Figure \ref{fig:sorting} (left column), observed sorting of lipids within {5-20~nm} of the \nachr~is dependent on the overall composition of the membrane. For all compositions shown, cholesterol is depleted within 5-20~nm and enriched even farther from the protein.  Within the binary systems this effect is minor ( $\tilde\rho_{CHOL} \sim 1$), but it becomes stronger in the moderately demixed LA systems ($\tilde\rho_{CHOL} \sim 0.5$) and substantial ($\tilde\rho_{CHOL} \sim 0.25$) for the highly-segregated DHA containing systems.  A similar pattern is observed for DPPC, which suggests that ``sorting'' over the 5-20~nm range is primarily driven by intrinsic differences in membrane organization that would be observed without the receptor. PUFAs are also most highly enriched at intermediate distances : the deepest red band is found at about 5~nm~in LA-containing systems and about 8~nm~ in DHA-containing systems.  This would be expected when \nachr~partitions near a curved domain boundary, as in Figure \ref{fig:SIQ}.      			

 \section{Discussion}
\label{S:4}

In this work we used coarse-grained simulations to predict the local lipid composition around the nicotinic acetylcholine receptor, in a range of domain forming membranes.  We observed  \nachr~partitioning to the liquid-disordered phase in all systems for which such a phase was present.  This is inconsistent with the model of lipid rafts as platforms that contain a high density of nAChRs, and unexpected in light of the established cholesterol dependence of nAChR.  As shown in these simulations, partitioning to the $\ldo$ phase does not prevent \nachr~from accessing cholesterol. 

The simulations presented here involve only one receptor per system.  Using the present results only, the simplest extrapolation to multiple receptors would assume that receptors are simply distributed randomly across the $\ldo$ domain.  The local receptor area density would be the number of receptors divided by the total area of $\ldo$ domains.    

In the model membranes used here, as well as in native nAChR membranes, the lipid composition would be expected to yield $\ldo$ phases that were about the same size as $\lo$ phases. The $\lo$ ``raft'' in an $\ldo$ ``sea'' analogy is not representative when over 50\% of the membrane is in the ``raft'' phase.  A more representative analogy would be receptors as boats, floating on an $\ldo$ lake within an $\lo$ rigid land mass.  Filling in the lake by adding to the coastline would force any boats in the lake closer together.  Similarly, any process that decreased total $\ldo$ area while keeping the number of receptors constant would increase the local receptor density.   In this model, observing increased \nachr~density by adding membrane cholesterol (as in \cite{
Barrantes2014,Bruses2001,Marchand2002,Oshikawa2003,Pato2008,Zhu2006, Barrantes2007,Wenz2005,Borroni2016}) would be consistent with \nachr~partitioning to the cholesterol-poor phase rather than the cholesterol-rich phase. 

This extrapolation from a single receptor assumes that introduction of additional receptors does not change partitioning behavior.  We do still find reliable partitioning to the $\ldo$ phase upon adding more receptors, and we will characterize systems with multiple receptors in a future publication. Due to receptor dimerization and trimerization, distribution of individual receptors within the $\ldo$ phase will not be random.   This would not change the expected trend of density increasing with added cholesterol, however.  

Observed partitioning into the $\ldo$ phase could be considered inconsistent with interpretations of some experiments, \cite{Bermdez_Partition_2010,Perillo_Transbilayer_2016}	which suggest minimal nAChR partitioning preference in symmetric model membranes or an actual preference for an $\lo$ phase in asymmetric model membranes.  These experiments used only monounsaturated acyl chains, and may have had less well-defined domains.  They further relied on detergent resistant membrane (DRM) methods, which are sensitive to {the} choice of detergent \cite{Brown2007} and could be unable to distinguish between proteins with no partitioning preference vs proteins that persistently partition to one side of a boundary. 

The origin of preferential partitioning observed in these simulations for the $\ldo$ domain is still unclear, but may reflect different elastic properties of the $\ldo$ and $\lo$ domains.  In general, proteins embedded in membranes will introduce a boundary condition on the membrane shape, such that (1) the thickness of the membrane matches the thickness of the transmembrane domain\cite{Aranda-Espinoza1996, Jensen2004, Brannigan2006} and (2) interfacial lipids are parallel to the protein surface.\cite{goulian1993}.  Transmembrane proteins with hydrophobic mismatch with the surrounding membrane may deform the membrane thickness to satisfy constraint (1), while cone-shaped proteins like pLGICs~ must also introduce a ``tilt'' deformation to satisfy (2).  Each leaflet of the membrane has an elastic resistance to bending away from its spontaneous curvature, and satisfying these constraints is energetically costly.  

Continuum theories based on the Helfrich Hamiltonian have been used to predict shape deformations around protein inclusions in homogeneous membranes.\cite{goulian1993,Aranda-Espinoza1996,Brannigan2006}  In mixed membranes, minimization of the protein-deformation free energy may also induce lipid sorting.  Two distinct sorting mechanisms could minimize the bending free energy: sorting that A) reduces the required bending deformation, by selecting boundary lipids with a specific thickness, leaflet asymmetry, or shape or B) reduces the free energy cost of the bending deformation, by selecting for flexible boundary lipids.   Mechanism (B) is the most generally applicable approach, and would stabilize partitioning to the most flexible domains, consistent with our observations (Figure S2).  In some cases, mechanism (A) may also contribute to partitioning or lipid-sorting, and could explain why \nachr~tends to attract saturated PE over saturated PC, or how leaflet asymmetry can promote partitioning to more rigid phases as observed in \cite{Perillo_Transbilayer_2016} . 

We previously \cite{Brannigan_Embedded_2008} proposed unresolved density in the cryo-EM structure of \nachr~ in the {\it Torpedo} membrane could be embedded cholesterol, based on gain of function caused by cholesterol in reconstitution mixtures\cite{Fong_Correlation_1986,Sunshine_Lipid_1992,Hamouda_Assessing_2006,Butler_FTIR_1993,Bhushan_Correlation_1993,Fong_Stabilization_1987,Bednarczyk_Transmembrane_2002,Corrie_Lipid_2002}, but we did not consider the possibility of occupation by polyunsaturated chains.  
Here we observe spontaneous binding of cholesterol to coarse-grained embedded sites, but long-chain PUFA tails displace cholesterol in some binding sites. Long acyl chains may penetrate far into the TMD bundle without requiring the entire head group also be incorporated, and long-chain PUFAs may do so without as substantial an entropic penalty as long saturated chains.  Cholesterol (like phosphatidic acid, another lipid known to cause gain of function under some preparations\cite{Butler_FTIR_1993,Bhushan_Correlation_1993,Fong_Stabilization_1987,Bednarczyk_Transmembrane_2002,Corrie_Lipid_2002})  has a much smaller headgroup than PC or PE. It can become fully incorporated into the TMD without the TMD needing to accommodate the bulky headgroup.   These complex associations underlie the challenges of predicting local lipid environment in heterogeneous, highly non-ideal mixtures. 

All simulations reported here contain lipids with di-saturated tails or di-PUFA tails. While lipid species with two identical acyl chains do exist in the native membrane, they are far less common than hybrid lipids with heterogeneous acyl chains.  Including hybrid lipids would reduce the potential for formation of large domains, while increasing the length of the domain interface. 
Incorporation of hybrid lipids would also reduce the \nachr-local concentration of PUFA chains.  Even 5-10\% DHA is a saturating concentration for \nachr~cavities, however, so we expect occupation of cavities to be minimally affected by replacement of di-DHA lipids with twice the number of hybrid lipids.

 \section{Acknowledgment}
GB and RS were supported by research grants NSF MCB1330728 and NIH P01GM55876-14A1. GB and LM were also supported through a grant from the Research Corporation for Scientific Advancement. This project was supported with computational resources from the National Science Foundation XSEDE program through allocation NSF-MCB110149, a local cluster funded by NSF-DBI1126052, the Rutgers University Office of Advanced Research Computing (OARC) and the Rutgers Discovery Informatics Institute (RDI2), which is supported by Rutgers and the State of New Jersey. We are grateful to Dr. J\'{e}r\^{o}me H\'{e}nin for his helpful suggestions throughout this study.

\bibliographystyle{model1-num-names}
\bibliography{readcube_export,CysLoopReferencesJuly2018}
\end{document}